\documentclass[aps,preprint]{revtex4}%
\usepackage{amsfonts}
\usepackage{amsmath}
\usepackage{amssymb}
\usepackage{graphicx}%
\providecommand{\U}[1]{\protect\rule{.1in}{.1in}}
\begin{document}
\title[ ]{Energy loss of charged particles in a two-dimensional Dirac plasma}
\preprint{ }
\author{Aqsa Arshad and Kashif Sabeeh$^{\dagger}$}
\affiliation{Department of Physics,Quaid-i-Azam University, Islamabad 45320, Pakistan}
\author{M. Tahir}
\affiliation{Department of Physics, University of Sargodha, Sargodha 40100, Pakistan}
\author{}
\affiliation{}
\pacs{PACS number}

\begin{abstract}
The stopping power and energy loss rate of charged particles traversing a
two-dimensional Dirac plasma is investigated. The Dirac plasma considered here
models a solid state system, recently realized graphene monolayer, where the
conduction electrons obey the Dirac-like equation and exhibit a linear in
momentum dispersion relation. Theoretical work presented here is based on the
the dielectric response function and the dynamical structure function within
the Random-Phase-Approximation (RPA).

\end{abstract}
\volumeyear{year}
\volumenumber{number}
\issuenumber{number}
\eid{identifier}
\date[Date text]{date}
\received[Received text]{date}

\revised[Revised text]{date}

\accepted[Accepted text]{date}

\published[Published text]{date}

\startpage{1}
\endpage{2}
\maketitle

\section{INTRODUCTION\textbf{ }}

Recent fabrication of single layers of graphite (graphene monolayer)\cite{1}
has posed the interesting question of charged particle energy loss in solid
media where the conduction electrons behave distinctly different from those
found in ordinary semiconductor heterostructures. In a two-dimensional
electron gas (2DEG) realized in semiconductor heterostructures, the conduction
electrons behave as ordinary electrons with a parabolic dispersion relation.
In contrast, the dispersion relation obeyed by conduction electrons in
graphene monolayer is linear in momentum. This occurs due to the unique
crystal structure of graphene which is a two-dimensional (2D) honeycomb
lattice of carbon atoms. Quantum-mechanical hopping between the sublattices of
graphene leads to the formation of \ two energy bands, and their intersection
near the edges of the Brillioun zone yields the conical energy spectrum. As a
consequence, the dispersion relation of electrons and holes bands is linear
near $K,\ K^{\prime}$ points of the Brillioun zone which is given by
$\epsilon_{k}=\hbar v_{F}k.$ Hence, the conduction electrons, known as Dirac
electrons, behave as massless particles with the effective speed of light
$v_{F}\simeq10^{6}m/s\approx c/300$\cite{2}. In a graphene monolayer, electron
transport is essentially governed by Dirac's (relativistic) equation, rather
than the usual Schrodinger equation for nonrelativistic quantum particles. The
relativistic behavior of graphene was first predicted by P. R. Wallace\cite{3}%
. It thus provides a unique opportunity to study relativistic quantum dynamics
in condensed-matter systems. The Dirac-like gapless energy spectrum was also
confirmed recently by cyclotron resonance measurements in graphene
monolayer\cite{4}.

In this work, we address energy loss experienced by charged particles as they
traverse a non-local dynamic quantum plasma of Dirac electrons, electrons that
obey the Dirac equation. The energy loss calculation performed here requires
that we determine the dynamic, non-local dielectric response function for the
plasma of Dirac electrons. We determine the dielectric response function
within the Random-Phase-Approximation(RPA) and employ it to calculate the
dynamical structure function which yields the stopping power.

Since the work of Nozieres and Pines\cite{5}, conduction electrons in solids
have been treated as a solid state plasma, a gas of electrons in a
neutralizing positive background. Moreover, electron energy loss spectroscopy
has been an important probe of the dielectric response properties of solid
state systems\cite{6}. This was brought into clear focus by the pioneering
work of Ritchie\cite{7} where the energy loss of a charged particle traversing
a bounded solid state plasma was considered. Dynamical processes in a solid
state medium have been investigated with a variety of spectroscopic tools
which employ an almost monochromatic beam of electrons, photons, neutral atoms
or ions which scatter inelastically from the medium under study\cite{8}.
Information concerning the electronic properties of the medium is obtained as
a result of the interaction between the probes and the elementary excitations
of the medium. Experimentally, this information is obtained by electron energy
loss spectroscopy (EELS) and inelastic light scattering studies. Recently,
energy loss spectroscopy of free-standing graphene films was performed\cite{9}

The present paper is arranged as follows. In section II we give the
formulation of the problem. The energy loss rate and stopping power of a 2D
Dirac plasma is outlined. This theoretical investigation is based on the
evaluation of dynamical structure function and energy loss function. In
section III, we present the results and discussion. Concluding remarks are
made in section IV.

\section{FORMULATION}

The system under consideration is a 2D quantum plasma of Dirac electrons i.e.
Dirac electrons embedded in a uniform and rigid neutralizing background of
positive charges. We begin by considering the low-energy electronic
eigenstates for this system, given by%
\begin{equation}
H_{0}\mathbf{F(r)}=\varepsilon\mathbf{F(r).} \label{1}%
\end{equation}
The Dirac-like Hamiltonian in eq(1) is\cite{10}%
\begin{equation}
H_{o}=v_{F}%
\begin{pmatrix}
0 & \hat{k}_{x}-i\hat{k}_{y}\\
\hat{k}_{x}+i\hat{k}_{y} & 0
\end{pmatrix}
=v_{F}(\sigma_{x}\hat{k}_{x}+\sigma_{y}\hat{k}_{y}) \label{2a}%
\end{equation}
where$\ \sigma_{x},\sigma_{y}$ are Pauli matrices and $v_{F}$\ is the
two-dimensional Fermi velocity of Dirac electrons (with the characteristic
velocity $v_{F}\simeq10^{6}m/s)$\cite{10}. Conduction electrons in a Dirac
plasma behave as massless Dirac particles with a linear dispersion relation
$\epsilon_{sk}=skv_{F}$ \cite{1} where $s=\pm1$ indicate the conduction $(+1)$
and valence $(-1)$ bands, respectively.\ $\mathbf{k}$ is the two-dimensional
wave vector and is given by $\mathbf{k}=k_{x}\hat{\imath}+k_{y}\hat{\jmath}$,
here $k_{x}=k\cos\theta_{k},\ k_{y}=k\sin\theta_{k}.$ Hence, electron dynamics
is modelled by the following Dirac equation
\begin{equation}
v_{F}(\sigma_{x}k_{x}+\sigma_{y}k_{y})\mathbf{F}_{sk}\mathbf{(r)}%
=\varepsilon\mathbf{F}_{sk}\mathbf{(r)}. \label{3}%
\end{equation}
From here onwards we take $\hbar=c=1$ throughout this calculation. The
wavefunctions appearing in eq.(\ref{3}) are given by\cite{10}%
\begin{equation}
\mathbf{\mathbf{\mathbf{F}_{s,k}\mathbf{(r)=}}}\frac{1}{L}%
\mathbf{\mathbf{\mathbf{F}_{s,k}\exp(i\mathbf{k.}}r}), \label{4}%
\end{equation}%
\begin{equation}
\mathbf{F}_{s^{\prime},k+q}\mathbf{(r)}=\frac{1}{L}\mathbf{F}_{s^{\prime}%
,k+q}\exp(i(\mathbf{k+q).r}) \label{5}%
\end{equation}
where $L^{2}$ is the area of the plasma sheet and%
\begin{equation}
\mathbf{F}_{sk}=\frac{1}{\sqrt{2}}\binom{e^{-i\theta_{k}}}{s} \label{6}%
\end{equation}
and%
\begin{equation}
\mathbf{F}_{s^{\prime},k+q}\mathbf{=}\frac{1}{\sqrt{2}}\binom{e^{-i\theta
_{k+q}}}{s^{\prime}}. \label{7}%
\end{equation}
The wavevector and frequency dependent longitudinal dielectric response
function within the Random-Phase-Approximation(RPA) can be expressed
as\cite{7}%
\begin{equation}
\epsilon(\mathbf{q},\omega)=1-\nu_{c}{\Large \pi}(\mathbf{q},\omega) \label{8}%
\end{equation}
where $\nu_{c}=2\pi e^{2}/\kappa q$ is the Fourier transform of two
dimensional coulomb interaction, $\kappa$ is the background dielectric
constant and ${\Large \pi}(\mathbf{q},\omega)\ $is the two-dimensional(2D)
polarizability. The plasmon modes at finite wave vectors are given by the
zeroes of the dielectric response function given by eq(\ref{8}).\ Since, in
this work, we are primarily interested in the energy loss due to plasmons, we
consider the dielectric response function in the high frequency, long
wavelength limit $\left(  q\longrightarrow0\right)  ,$\ which is given as
\cite{11}%
\begin{equation}
\epsilon(\mathbf{q},\omega)\approx1-\frac{\omega_{p}^{2}}{\omega^{2}}\left(
1-\frac{\omega^{2}}{4E_{F}^{2}}\right)  \label{9a}%
\end{equation}
Here $\omega_{p}=$ $\left(  g_{s}g_{v}e^{2}E_{F}/2\kappa\right)  ^{1/2}%
\sqrt{q}\ $is the plasma frequency with $E_{F}$ the Fermi energy.
Equivalently, in terms of dimensionless variables, we have%
\begin{equation}
\epsilon(x,\nu)\approx1-\frac{2\pi r_{s}}{k_{F}}\left(  \frac{g_{s}g_{v}n_{e}%
}{\pi}\right)  ^{1/2}\frac{x}{2\nu^{2}}\left(  1-\frac{\nu^{2}}{2}\right)  .
\label{10}%
\end{equation}
In the above expression, we have used the dimensionless variables
$x=q/k_{F},\ \nu=\omega/E_{F},$.$r_{s}\ $is the Wigner-Seitz radius,$\ \kappa$
is the background dielectric constant and $g_{s}=$ $g_{v}=2$ being the spin
and valley degeneracies respectively\cite{11}. Energy loss function is the
basic parameter which accounts for the energy lost by the incident charged
particle. It is also the parameter which is of central importance in energy
loss spectroscopy (EELS) experiments. Energy loss function is defined as the
imaginary part of the inverse dielectric response function. From the
dielectric response function given by eq(\ref{9a}) and introducing the
infinitesimally small parameter $\eta$ we obtain%
\begin{equation}
\frac{1}{\epsilon(q,\omega)}=\frac{(\omega+i\eta)^{2}}{(\omega+i\eta
)^{2}-\omega_{p}^{2}\left(  1-\frac{(\omega+i\eta)^{2}}{4E_{F}^{2}}\right)
}=\frac{A}{\omega-\omega_{p}\sqrt{1-\frac{\omega^{2}}{4E_{F}^{2}}}+i\eta
}+\frac{B}{\omega+\omega_{p}\sqrt{1-\frac{\omega^{2}}{4E_{F}^{2}}}+i\eta}
\label{10a}%
\end{equation}
Evaluating the coefficients $A,\ B$ and using Dirac's prescription $\lim
_{\eta\longrightarrow0}\frac{1}{\xi\pm i\eta}=\frac{1}{\xi}\mp i\pi\delta
(\xi)$, in eq(\ref{10a}) and then expressing the variables in terms of
dimensionless parameters $x$ $\nu$ and $r_{s}$ we obtain the energy loss
function for graphene monolayer as
\begin{equation}
\operatorname{Im}\left(  \frac{1}{\epsilon(x,\nu)}\right)  =\frac{-\pi\left(
\frac{g_{s}g_{v}r_{s}x}{8}\right)  ^{1/2}}{\left(  1-\frac{\nu^{2}}{4}\right)
}\left[
\begin{array}
[c]{c}%
\delta\left(  \frac{\nu}{\left(  1-\frac{\nu^{2}}{4}\right)  }-\sqrt
{\frac{g_{s}g_{v}r_{s}x}{2}}\right) \\
-\delta\left(  \frac{\nu}{\left(  1-\frac{\nu^{2}}{4}\right)  }+\sqrt
{\frac{g_{s}g_{v}r_{s}x}{2}}\right)
\end{array}
\right]  . \label{11}%
\end{equation}
The interaction of charged particles with condensed matter system can be
studied by means of the system's stopping power. The energy loss per unit path
length is the stopping power $S\equiv-dE/dx$ and it accounts for the energy
lost by an external charged projectile as it passes through matter. In the
quantum mechanical framework, we consider an in-plane probe of charge $ze,$
mass $m$ and velocity $v_{p},$ interacting with the many particle system under
consideration by treating the incident particle state as a plane wave state.
We calculate the probability that the point-like projectile loses an energy
$\omega$ in a time interval $dt$ in interaction with the 2D Dirac plasma.
During the interaction impulse transfer is $q.$ Here $\omega$ and $q$ satisfy
$\omega=\mathbf{q.v}_{p}+q^{2}/2m.$ For a heavy incident particle, we neglect
the recoil energy $q^{2}/2m$ to obtain $\omega=\mathbf{q.v}_{p}.$The stopping
power is given by summing over energy difference weighted by transition rate
$W$ times the inverse projectile velocity, mathematically$\ $%
\begin{equation}
dE/dx\equiv-1/v_{p}%
%TCIMACRO{\dsum \limits_{\mathbf{q}}}%
%BeginExpansion
{\displaystyle\sum\limits_{\mathbf{q}}}
%EndExpansion
(\varepsilon_{\mathbf{k}}-\varepsilon_{\mathbf{k\ -q}})W. \label{11a}%
\end{equation}
We can expand energy loss in terms of $\Delta\omega\equiv$ $q^{2}/2m$ to
obtain \cite{12}%
\begin{equation}
\frac{dE}{dx}=\left(  \frac{dE}{dx}\right)  _{0}+\left(  \frac{dE}{dx}\right)
_{1}+...... \label{11b}%
\end{equation}
Here we also make use of the sum rule $-%
%TCIMACRO{\dint \limits_{0}^{\infty}}%
%BeginExpansion
{\displaystyle\int\limits_{0}^{\infty}}
%EndExpansion
\omega\operatorname{Im}\left\{  1/\epsilon(q,\omega)\right\}  d\omega
=\pi\omega_{p}^{2}/2$. To the lowest order, the stopping power for a 2D Dirac
plasma given by eq(\ref{11a}) can be expressed in terms of the dynamical
structure functions $S(\mathbf{q,}\omega)$ as%
\begin{equation}
\left(  \frac{dE}{dx}\right)  _{0}=-\frac{1}{v_{p}}\int_{0}^{\infty}%
%TCIMACRO{\dint _{0}^{qv_{p}}}%
%BeginExpansion
{\displaystyle\int_{0}^{qv_{p}}}
%EndExpansion
d\omega d^{2}\mathbf{q}\left(  \frac{n_{e}^{2}e^{2}\omega^{2}}{\kappa
q}\right)  ^{2}S(\mathbf{q,}\omega)\ \delta(\omega-\mathbf{q.v}_{p})
\label{11.1}%
\end{equation}
whereas the first order stopping power is given by%
\begin{equation}
\left(  \frac{dE}{dx}\right)  _{1}=-\frac{1}{2mv_{p}}\int\int\frac{n_{e}%
q^{2}d^{2}\mathbf{q}}{(2\pi)^{2}}\left(  \frac{2\pi e^{2}}{\kappa q}\right)
^{2}\frac{\partial}{\partial\omega}\left[  \omega S(\mathbf{q,}\omega
)\ \delta(\omega-\mathbf{q.v}_{p})\right]  d\omega. \label{11.1.2}%
\end{equation}
The dynamical structure function is the auto-correlation function of the
Fourier components of the particle density and accounts for longitudinal
charge oscillations of the electron density relative to the positive
background. From the polarizability, we can obtain the dynamical structure
factor at finite temperature for a 2D Dirac plasma as%
\begin{equation}
S(\mathbf{q,}\omega)=\frac{\kappa^{2}\omega_{p}^{2}}{2g_{s}g_{v}e^{4}%
n_{e}E_{F}\left(  1-\frac{\omega^{2}}{4E_{F}^{2}}\right)  }(1-\exp
(-\beta\omega))^{-1}\left[  \delta(\omega_{0}-\omega_{p})-\delta(\omega
_{0}+\omega_{p})\right]  \label{11.2}%
\end{equation}
where $\omega_{0}=\omega/\sqrt{1-\omega^{2}/4E_{F}^{2}},\ \beta=1/k_{B}%
T,\ k_{B}$ is the Boltzmann constant, $T$ is the absolute temperature,
$E_{F}=v_{F}k_{F},\ k_{F}=(4\pi n_{e}/g_{s}g_{v})^{1/2}$ is the Fermi
wavevector and $n_{e}\ $is the electron density. Dynamical structure factor
provides direct physical information about longitudinal excitations in the
system. It is the measure of the density-density correlations of the system.
Calculations of dynamical structure function reveal that the electromagnetic
response of a Dirac plasma is substantially different from a 2DEG system which
is essentially due to the distinctly different dispersion relation of Dirac
electrons in a 2D Dirac plasma as compared to ordinary electrons in a 2DEG system.

For a solid state Dirac plasma under consideration, we can classify the rate
of energy loss and stopping power into two categories, one due to the
particle-hole excitations and the other due to the plasma oscillations. So the
total stopping power of the system is%
\begin{equation}
\frac{dE}{dx}=\frac{dE_{e-h}}{dx}+\frac{dE_{pl}}{dx} \label{12}%
\end{equation}
Employing the following relation in eq(\ref{11.1})
\begin{align}
&  \left.  \delta(\varepsilon-\varepsilon_{n}-V_{n}\cos(\frac{2\pi}{a}%
x_{0}))=\frac{a}{2\pi}\left[  \delta(x_{0}-\frac{a}{2\pi}\arccos
(\frac{\varepsilon-\varepsilon_{n}}{V_{n}}))+\delta(x_{0}+\frac{a}{2\pi
}\arccos(\frac{\varepsilon-\varepsilon_{n}}{V_{n}}))\right]  \right.
\nonumber\\
&  \left.  \times\frac{1}{\sqrt{V_{n}^{2}-(\varepsilon-\varepsilon_{n})^{2}}%
}\Theta(\left\vert V_{n}\right\vert -\left\vert \varepsilon-\varepsilon
_{n}\right\vert )\right.  \label{12.1}%
\end{align}
For a single charged particle$,\ z=1,$ the lowest order stopping power, due to
plasma oscillations, of a 2D degenerate Dirac plasma is given by%
\begin{equation}
S\approx\frac{dE_{pl}}{dx}=-\frac{1}{v_{p}}\int_{0}^{q_{c}}\frac{e^{2}%
\omega_{p}^{2}}{2\kappa\sqrt{1-\frac{\omega_{p}^{2}}{4E_{F}^{2}}}}\frac
{1}{\sqrt{(qv_{p})^{2}-\left(  \omega_{p}\sqrt{1-\frac{\omega_{p}^{2}}%
{4E_{F}^{2}}}\right)  ^{2}}}\ln\left(  y+\sqrt{y^{2}-1}\right)  \arccos h(y)dq
\label{12.2}%
\end{equation}
with $y=qv_{p}/\omega_{p}\sqrt{1-\frac{\omega_{p}^{2}}{4E_{F}^{2}}}$

\section{RESULTS AND DISCUSSION}

The dynamic and static response properties of an electron system are all
embodied in the structure of the dielectric response function. We have
employed the Random-Phase- Approximation (RPA) based dielectric response
function to determine the structure function and the energy loss function of a
2D Dirac plasma. The two types of excitations in the medium responsible for
energy loss are the collective and single particle excitations. Collective
excitations (plasmons) occur at a higher energy compared to single particle
(electron-hole) excitations. At a frequency $\omega_{0}$ equal to $\omega
_{p},$ collective excitations (plasma oscillations) dominate for $q^{2}%
v_{p}^{2}<\omega_{p}^{2}$ and these exist as long as the wavevector remains
less then a critical value $q_{c}.\ $However, above this critical value
plasmons undergo Landau damping generating electron-hole pair excitations.
Therefore, energy loss rate and stopping power are affected by these two types
of contributions, one due to the collective oscillations i.e. plasmons and the
other due to the electron-hole excitations. The energy loss function taking
into account both these mechanisms of energy loss is presented here.
Collective modes with the onset of Landau damping change into electron-hole
pairs. We have plotted the loss function given by eq(\ref{11}) due to
collective excitations in Fig.(1). Here we essentially restrict the energy
loss to plasmon excitations by plotting the result at
$x=0.1,\ 0.5,\ 0.01,\ 0.05,\ 0.001.$The parameters used are: electron density
$n_{e}=3.16\times10^{15}/m^{2}$, $v_{F}=10^{6}m/\sec,$ $\kappa=2.5$(using
SiO$_{2}$ as the substrate material), $r_{s}=0.5$. Note that the critical
value of the wave vector $q_{c}$ and the corresponding critical value of $x$
is $0.7$ where the onset of Landau damping of plasmons occur for the
parameters considered here. Dynamical structure function shows delta function
peaks at the plasma frequency. The peaks in the figure are the manifestation
of existence of plasmons in the system. In Fig.(2), in order to include the
contribution of single particle excitations, we have plotted energy loss
function against $\nu=\omega/E_{F}$ for following values of $x\equiv
q/k_{F}=1.0,\ 1.5,\ 2.0.$ The peaks in the energy loss function seen in the
figure can be interpreted as occurring due to single particle excitations of
the system. In Fig.(3), we have plotted energy loss function versus
$x=q/k_{F}$ for various values of $\nu=\omega/E_{F}.$ The relationship between
energy loss rate $dE_{pl}/dt$ due to plasmons and velocity of the incident
particle $v_{p}$ is shown graphically in figure(4). We have plotted the energy
loss rate versus the dimensionless velocity $v_{p}/v_{F}$. Stopping power is
shown in Fig.(5) where the energy lost by the incident particle is due to
plasma oscillations. The stopping power decreases with the increase of the
incident particle velocity. Furthermore, the lowest order energy loss rate and
stopping power of the medium are found to be independent of the mass of the
incident particle.

\section{CONCLUSIONS}

An analysis of energy loss suffered by fast particles in their interaction
with matter requires the calculation of energy loss rate and stopping power.
In this work, we present a theoretical investigation of energy loss through
the calculation of both the energy loss rate and the stopping power of a 2D
solid state Dirac plasma. The motivation of this work is the recent
realization of single layer of graphite, graphene monolayer. The
quasiparticles in graphene monolayer are found to obey the Dirac equation with
a linear in momentum dispersion relation. The work presented here is based on
the the dielectric response function and the dynamical structure function
within the Random-Phase-Approximation (RPA). The energy loss function taking
into account the collective as well as single particle mechanisms of energy
loss is presented here. In the collective excitation regime, energy loss
function peaks at the plasma frequency, such peaks when detected by
experiments, can also be used to identify energy at which plasmons occur in
the medium. Furthermore, stopping power as a function of the incident particle
velocity is also determined. Stopping power is found to decrease as the
velocity of the incident particle is increased. Moreover, the lowest order
energy loss rate and stopping power of the Dirac plasma are found to be
independent of the mass of the incident particle.

\section{Acknowledgements}

K. Sabeeh) would like to acknowledge the support of the Pakistan Science
Foundation (PSF) through project No. C-QU/Phys (129).

$\dag$Electronic address: ksabeeh@qau.edu.pk; kashifsabeeh@hotmail.com

\end{document}